\documentclass[11pt,twoside]{article}

\usepackage{asp2004}
\usepackage{epsf}

\begin{document}
\title{Properties of the Scattering Screen in Front of Scintillating Quasar PKS 1257-326}   
\author{H.~E.~Bignall}
\affil{Joint Institute for VLBI in Europe, Postbus 2, 7990AA Dwingeloo, The Netherlands}    
\author{D.~L.~Jauncey, J.~E.~J.~Lovell, A.~K.~Tzioumis}
\affil{Australia Telescope National Facility, CSIRO, PO Box 76, Epping, NSW 1710, Australia}
\author{J.-P.~Macquart}
\affil{Jansky Fellow at NRAO/Caltech. Department of Astronomy, Mail Code 105-24 Robinson, Caltech, Pasadena CA 91125, USA}
\author{L.~Kedziora-Chudczer}
\affil{Institute of Astronomy, School of Physics A28, The University of Sydney, NSW 2006, Australia}

\begin{abstract} 
PKS 1257--326 is a quasar showing extremely unusual, rapid
interstellar scintillation (ISS), which has persisted for at least a
decade.  Simultaneous observations with the VLA and ATCA, combined
with ATCA monitoring over several years, have revealed some properties
of the turbulent ionized medium responsible for the ISS of PKS
1257--326. The scattering occurs in an unusually nearby ($\sim
10$\,pc), localized ``screen''.  The scintillation pattern is highly
anisotropic with axial ratio more than $10:1$ elongated in a northwest
direction on the sky. Recent findings and implications for small-scale
ionized structures in the ISM are discussed.
\end{abstract}


\section{Introduction}   
Interstellar scintillation (ISS) of compact radio sources probes
structure in the ionized instellar medium of our Galaxy down to
extremely small scales ($\sim 10^5$\,km).  PKS~1257$-$326
(Galactic coordinates $l=305.2^{\circ}, b=29.9^{\circ}$)  is a flat
spectrum, radio-loud quasar at $z=1.256$ which exhibits intrahour
flux density variability at frequencies of several GHz due to ISS
\citep{big2003}. It was the third confirmed ``intrahour variable''
(IHV) quasar, after PKS~0405$-$385 \citep{ked97}
and J1819+3845 
\citep{dtdb2000}.

The recent large-scale {\it MicroArcsecond Scintillation-Induced
Variability} (MASIV) VLA Survey at 5~GHz showed that while a large
fraction of flat-spectrum radio sources exhibit variations of
typically a few percent on timescales of order a day, the IHV
phenonemon is extremely rare \citep[][and these
proceedings]{lov2003}. Only a handful of quasars are known to show
large variations on timescales of a few hours or less.  While a source
must be compact in order to scintillate, short scintillation
time-scales have been attributed to nearby scattering screens within a
few tens of parsec from the Sun \citep{dtdb2000,ric2002}.

\section{Observations and analysis}
\begin{figure}
\plotfiddle{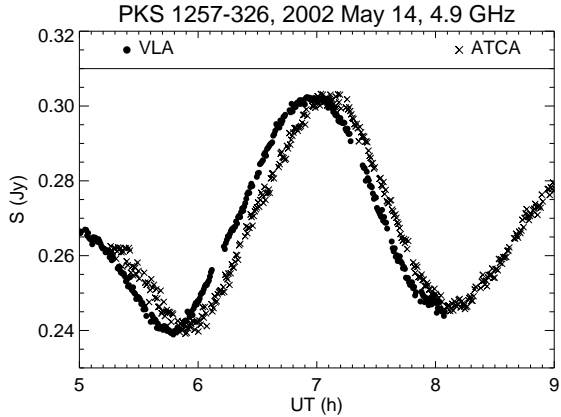}{55mm}{0}{45}{45}{-140}{0}
\caption{Simultaneous flux density measurements of PKS~1257$-$326 at
the VLA and ATCA on 2002 May 14 at 4.9\,GHz. The pattern takes 8
minutes to drift over the $10^4$~km baseline, yet the annual
cycle in timescale implies a scintillation velocity significantly larger than
20\,$km\,s^{-1}$ at this time of year. The large observed time delays in May
are due to a highly anisotropic pattern elongated almost perpendicular to the
baseline and moving with a significant velocity component parallel to
the baseline.}
\label{fig-delay}
\end{figure}

We began monitoring PKS~1257$-$326 with the ATCA at 4.8 and 8.6~GHz in
2000 following the discovery of rapid variability \citep{big2003}.  We
found an annual cycle in the timescale of variability which is
evidently consistent over several years of observations at both
frequencies. The annual cycle results from the change in scintillation
velocity due to the Earth's orbital motion. Furthermore, simultaneous
observations of PKS~1257$-$326 with the VLA and ATCA revealed large
delays of up to 8 minutes between the variability pattern arrival
times at each telescope \citep{big2006}, as shown in
Figure~\ref{fig-delay}. The two-station time delays were measured in
three separate epochs, simultaneously at two frequencies and on two
consecutive days in each epoch. Unlike for the case of J1819+3845
observed with the VLA and WSRT \citep{dtdb2002}, where the projected
baseline rotated through a large angle during the simultaneous
observation and the time delay changed sign, for PKS~1257$-$326
observed with the VLA and ATCA we detect no significant change in
delay over the course of an observation. The mean pattern arrival time
delays for each epoch were $483 \pm 15$~s in 2002 May, $333 \pm 12$~s
in 2003 January, and $318 \pm 10$~s in 2003 March.   The observed
annual cycle and time delays not only demonstrate conclusively that
the rapid, large-amplitude variability of PKS~1257$-$326 is entirely
due to ISS, but moreover these results can be used quantitavely to
determine the bulk velocity of the scattering plasma as well as the
characteristic length scale of the scintillation pattern in two
dimensions.

In our analysis, the scintillation pattern is assumed to be elliptical
with minor axis characteristic scale length $a_{\rm min}$, and major
axis $a_{\rm maj} = R a_{\rm min}$ oriented along the vector $\hat{\bf
S}=(\cos \beta, \sin \beta)$.
The scintillation velocity is written in the form ${\bf v}(T)=v_{\rm
ISS} - v_\oplus(T) \equiv(v_\alpha (T),v_\delta (T))$, and varies on
an annual cycle as a function of time, $T$ \citep{mj2002}.  
Here we define $v_\oplus(T)$ as the
velocity of the Earth with respect to the solar system barycenter. 
In this notation, the time delay
expected between two telescopes displaced by a distance ${\bf r} =
(r_\alpha, r_\delta)$ is \citep{ck78}
\begin{eqnarray}
\Delta t = \frac{{\bf r} \cdot {\bf v} + (R^2 -1)({\bf r} \times
\hat{\bf S})({\bf v} \times \hat{\bf S} ) }{v^2 + (R^2-1)({\bf v}
\times \hat{\bf S})^2}, \label{TimeDelay}
\end{eqnarray}
where the dependence of ${\bf v}$ on $T$ is suppressed.  The
scintillation timescale is
\begin{eqnarray}
t_{\rm scint} = \frac{a_{\rm min}}{\sqrt{v^2 + (R^2 -1) ({\bf v}
\times \hat{\bf S})^2}}. \label{TScint}
\end{eqnarray}
We define $t_{\rm scint}$ as the point where the intensity
autocorrelation function decays to $1/e$ of its maximum value.

In practice neither the annual cycle nor the time delay
experiment datasets suffice to uniquely determine all five
scintillation parameters, $v_{{\rm ISS},\alpha}, v_{{\rm ISS},\delta},
\beta, R$ and $a_{\rm min}$, at any one observing frequency, so we fit to both
the time delay and the annual cycle data simultaneously. We also fit
both frequencies simultaneously, imposing the additional constraint
that the scintillation velocity is identical at the two frequencies;
$R$ and $\beta$ at the two frequencies are still assumed independent.

\subsection{Anisotropy}
When the scintillation pattern is highly anisotropic, there are
degenerate solutions for the axial ratio $R$ and the component of
velocity parallel to the major axis. However the characteristic
length scale along the minor axis, $a_{\rm min}$, can still be
uniquely determined.  In weak scattering, as observed for
PKS~1257$-$326 at frequencies of $\sim 5$~GHz and above, the minor
axis length scale is related to the Fresnel scale, $r_{\rm F} =
\sqrt{cL/(2\pi\nu)}$, where $\nu$ is the observing frequency and $L$
is the scattering screen distance.  Thus $a_{\rm min}$ can be used to
constrain $L$.

\begin{figure}
\plotfiddle{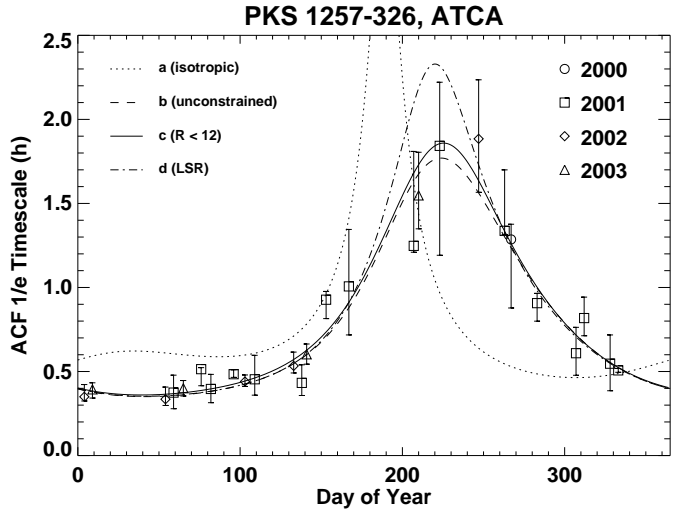}{65mm}{0}{55}{55}{-160}{-10}
\caption{Annual cycle in characteristic timescale, $t_{\rm scint}$,
  observed in ATCA data at 4.8\,GHz between 2000 and 2003. The plotted
  lines show model annual cycles for best fits to the combined time
  delay and annual cycle data with (a) the scintillation pattern fixed
  to be isotropic ($R=1$); (b) no constraints on the parameters; (c)
  pattern axial ratio limited to $R\leq 12$; (d) screen velocity fixed
  to the local standard of rest.} 
\label{fig-fit}
\end{figure}

For PKS~1257$-$326 we find that a highly anisotropic scintillation
pattern is required to fit the data. The position angle of anisotropy
is well constrained to within a few degrees but there is a degeneracy
between the two components of the scattering screen velocity, and the
pattern axial ratio $R$ is poorly constrained. Figure~\ref{fig-fit}
shows the timescale data and annual cycles from several
models imposing different constraints on the parameters. For the fits
shown, we weighted $\chi^2$ to give the (comparatively few) time delay
measurements equal importance in the fit relative to the
characteristic timescale measurements. The dotted line in
Figure~\ref{fig-fit} show that assuming an isotropic scintillation
pattern results in a very poor fit to the data.

\begin{figure}
\plotfiddle{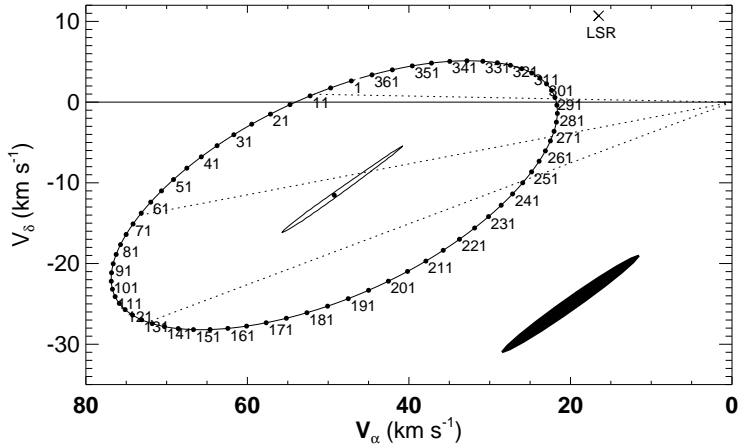}{50mm}{0}{60}{60}{-160}{-10}
\caption{Scintillation velocity projected onto the plane of the sky. The
large ellipse shows the annual variation of the Earth's velocity with
respect to a scattering screen moving at $-49.2$\,km\,s$^{-1}$ in RA and
$11.5$\,km\,s$^{-1}$ in Dec with respect to the Sun, corresponding to
solution (c) in Figure~\ref{fig-fit}. Dotted lines indicate the velocities
on the days of the time delay observations. The 1-$\sigma$ contour of the
best fit solution for $v_{{\rm ISS},\alpha}$ versus $v_{{\rm ISS},\delta}$
is plotted around the best fit screen velocity at the center of the large
ellipse. The component of velocity parallel to the long axis of the
scintillation pattern is poorly constrained. If the scattering screen were
moving with the local standard of rest, the large ellipse would be
centered at position $\times$ marked in the top right-hand corner. The
black filled ellipse represents the anisotropy (position angle and axial
ratio) of the best fit scintillation pattern at 4.9\,GHz.}
\label{fig-ellipse}
\end{figure}

Based on the similarity of the patterns observed at the two
telescopes, we can determine a lower limit for $R$. When the
scintillation velocity has a significant component perpendicular to
the baseline ${\bf r}$, as is the case for the May time delay data, we
can expect a small amount of spatial decorrelation of the
scintillation pattern over the ATCA-VLA projected baseline.  As
detailed in \citet{big2006} after  careful processing of the data we
can constrain the axial ratio $R$ to be at least $\sim 12:1$ based on
the degree of correlation between the patterns seen at each
telescope. Figure~\ref{fig-ellipse} shows  the annual cycle in
scintillation velocity for a scattering screen moving at
$-49.2$\,km\,s$^{-1}$ in RA and $11.5$\,km\,s$^{-1}$ in Dec with
respect to the Sun. The position angle of anisotropy in the
scintillation pattern ($-55^{\circ}$ North through East) is also
illustrated by the filled ellipse in Figure~\ref{fig-ellipse}.

As discussed by \citet[][and these proceedings]{rc2004}, there are
several lines of evidence for anisotropic turbulence in the ISM.
Moreover the intensity autocorrelation functions for
PKS~1257$-$326 display strong oscillatory behaviour which is a
signature of anisotropic scattering \citep{ric2002}. Therefore it
seems likely that the large anisotropy in the scintillation pattern
results from anisotropic scattering rather than anisotropic source
structure. Highly anisotropic scattering suggests a plasma
controlled by a magnetic field and localized in a relatively thin
``screen''.

\subsection{Distance to the scattering screen}
Our best fit models give $a_{\rm min} = 4.24 \pm 0.08 \times 10^4$~km
at 5~GHz, and $a_{\rm min} = 3.43 \pm 0.06 \times 10^4$~km at
8.6~GHz. For anisotropic scattering, our analysis shows that our
definition of $a_{\rm min}$ corresponds to $0.78 r_{\rm F}$ for a
point source in weak scattering. If the source angular size is larger
than the angular Fresnel scale then $a_{\rm min}$ will be larger than
$r_{\rm F}$. Thus our data put an upper limit on the screen distance
of $L \la 10$~pc.

\subsection{The influence of source structure?}
Although there is evidence for the scattering medium being largely
responsible for the high degree of anisotropy in the scintillation
pattern of PKS~1257$-$326, it is possible that source structure also
plays a role in the observed scintillation. 

If the scintillating component is circularly symmetric, then assuming
a screen distance of $10$\,pc, approximately 100\,mJy would have to be
contained within an angular size of $\sim 30~\mu$as in order to display the
observed large modulations. This implies a brightness temperature of
at least $T_b \sim 10^{13}$~K, requiring a Doppler factor of 10--20 to
reduce the intrinsic brightness temperature below the inverse Compton
limit. Alternatively, if the ``core'' happened to be elongated, or
consist of several components aligned in the same direction as the ISM
anisotropy, then this would reduce the brightness temperature
requirements. While an alignment of source structure with ISM
anisotropy seems an unlikely coincidence, it is conceivable that this
could be another ``selection effect'' for IHV.
Both VLA \citep{big2006} and VLBA \citep*[][in press]{jcam06} images of
the source show broad ``jet'' components extended to the northwest of
the core. Speculatively, if the unresolved VLBI core consists of
several components on sub-mas scales also aligned in this direction,
then this implies an approximate alignment of the jet with the major 
axis of anisotropy in the scintillation pattern. In this case, since
there is effectively only significant scattering in the direction
perpendicular to the jet, the scattering screen would not ``resolve''
structure along the jet axis.

\section{Conclusions}
Simultaneous observations with the VLA and ATCA, combined with ATCA
monitoring over several years, have revealed some properties of the
scattering plasma responsible for the rapid,
intrahour-timescale scintillation of PKS 1257--326. The scattering
occurs in an unusually nearby ($\sim 10$\,pc), localized
``screen''.  Based on the longevity
of the scintillation we can derive a lower limit of $10^{15}$~cm on
the linear extent of the screen.
Intrahour variability is extremely rare, suggesting that
these nearby screens cover a tiny fraction of the sky. The turbulence is
highly anisotropic, producing a scintillation pattern with axial ratio
more than $10:1$, elongated in a northwest direction on the sky.  It
was suggested by J.~Linksy (private communication, this conference) that
the scattering could possibly be produced in turbulent regions of
interaction at the edges of local clouds. It will be important to
investigate the possible ISM structures responsible for the rapid
scintillation of some quasars, not only to better understand the
tiny-scale structure in the ionized ISM, but also to gauge the effects
of ISS on observations with future sensitive cm-wavelength
telescopes.

\acknowledgements 
The ATCA is part of the Australia Telescope, which is funded by the
Commonwealth of Australia for operation as a National Facility managed
by CSIRO. The National Radio Astronomy Observatory is a facility of
the National Science Foundation operated under cooperative agreement
by Associated Universities, Inc.


\end{document}